\documentclass[11pt,a4paper]{article}
\usepackage{jheppub}

\title{Einstein Equations From Holographic Thermodynamics and Holographic
Entropy}

\author[a]{Miao Li,}

\author[b]{Rong-Xin Miao}

\author[a]{and Jun Meng}

\affiliation[a]{Kavli Institute for Theoretical Physics, National Key
Laboratory of Frontiers in Theoretical Physics, Institute of
Theoretical Physics, Chinese Academy of Sciences, \\
Beijing 100190, People's Republic of China.\footnote{mli@itp.ac.cn}
}

\affiliation[b]{Interdisciplinary Center for Theoretical Study, University of Science and Technology of China,\\
Hefei, Anhui 230026, People's Republic of
China.\footnote{mrx11@mail.ustc.edu.cn}}

\abstract{We derive the Einstein field equations and black hole
entropy from the first law of thermodynamics on a holographic
time-like screen. Because of the universality of gravity, the stress
tensor on the screen must be independent of the details of matter
fields, so it should be a pure geometric quantity. For simplicity,
we assume that the stress tensor on the screen depends on surface
Ricci curvature and extrinsic curvature linearly. Then we prove that
the surface stress tensor is just the Brown-York stress tensor plus
terms which do not affect the field equations of gravitation and the
entropy of the system. By assuming a generalized ``Fine first law of
thermodynamics" or the usual universal first law of thermodynamics
on the screen, we can derive the matter field equations as well.}

\begin{document}
\maketitle

\section{Introduction}

Ever since the discoveries of Hawking radiation \cite{Hawking} and
Black hole entropy \cite{Bekenstein}, the holographic viewpoint of
gravity and the relationship between gravity and thermodynamics have
draw much attention and brought about many research fields.

Along the line of holographic viewpoint of gravity, 't Hooft
\cite{Hooft} and Susskind \cite{Susskind} proposed the holographic
principle \cite{Bousso} for gravity theory, which states that the
description of a volume of space can be thought of as encoded on a
boundary to the region. The holographic principle was realized in an
exact example known as AdS/CFT by Maldacena \cite{Maldacena}, who
claims that the quantum gravity theory in $AdS_5\times S^5$
spacetime is dual to the $N=4$ Super-Yang-Mills theory without
gravity on the boundary. It should be mentioned that before
Maldacena, Brown and Henneaux had found that the asymptotic algebra
of 2+1 D gravity in asymptotic $AdS_3$ space is a Virasoro algebra
of 2 D CFT, which can be regarded as the first example of $AdS/CFT$
though the authors did not make such claims in their paper
\cite{Brown}.

Along another line of the connection between gravity and
thermodynamics, a noticeable breakthrough was made by Jacobson in
1995 \cite{Jacobson}, who first claims that gravity may not be a
fundamental interaction but emerges from the first law of
thermodynamics on a local Rindler horizon. This is a big claim on
the origin of gravity, although Jacobson only succeeded in deriving
the null-null component of Einstein equation. His work was then
generalized to the case of higher derivative gravity
\cite{Eling,Brustein}. Recently, another significant breakthrough
was made by Verlinde \cite{Verlinde}(see also Padmanabhan
\cite{Pad}), who states that the entropy will increase when
particles pass through a holographic screen, which causes the
emergence of gravity as an entropic force. Assuming the
equipartition rule and the Tolman-Komar mass, Verlinde succeeded to
obtain the time-time component of Einstein equation. There are many
advantages in Verlinde's approach, in particular, the holographic
screen can be placed at any place rather than only null surface in
Jacobson's approach. However, as pointed out in \cite{Li}, there is
negative-temperature problem when one locates the closed screen
arbitrarily. There have appeared many following up papers
\cite{reference} together with some criticism \cite{Kobakhidze}
after Verlinde's excellent work. However, it is still too early to
say whether the thermodynamic viewpoint of gravity is right or not.

In this paper, we combine the spirits of holographic and
thermodynamic viewpoints of gravity and prove that gravity can
emerge from the first law of thermodynamics on a time-like screen.
With the requirement that the entropy is a total differential, we
succeed in deriving all the components of Einstein equations and proving
that the entropy of a stationary black hole is the Bekenstein-Hawking
entropy. In our approach, the key hypothesis is that the stress
tensor on the screen depends on the surface Ricci curvature and
extrinsic curvature linearly, which is quite reasonable and natural.
Because gravity is universal, the surface stress tensor must be
independent of the details of matter fields, so it should be purely
geometrical. For simplicity, we assume that it depends on surface
Ricci curvature and extrinsic curvature only linearly. Then, we
prove that the stress tensor is just Brown-York stress tensor plus
terms irrelevant with the bulk field equations and entropy of the
system. It should be stressed that one may derive the higher
derivative gravity with a more general assumption of the surface
stress tensor. However, we only consider the simplest case in this
paper. Finally, we want to mention that we can also derive Einstein
equations form the entropy production of hydromechanics on the
holographic screen \cite{Miao}.

The paper is arranged as follows. In Sec.~2, we give a brief review
of thermodynamic viewpoint of gravity. In Sec.~3, we derive the
vacuum Einstein equations from the first law of thermodynamics on a
holographic time-like screen. In Sec.~4, we generalize our
holographic scheme to the cases of the Einstein equations with
matter fields. We conclude in Sec.~5.

\section{Brief review of thermodynamic viewpoints of gravity}

In this section, we shall briefly review the work of Verlinde
\cite{Verlinde} and Jacobson \cite{Jacobson}, and then compare their
approaches with ours of the next section.

First, let us review Verlinde's approach of the derivations of the
Einstein equation. According to \cite{Wald}, the Newton's potential
is
\begin{equation}\label{np}
\phi= \log(-\xi^{\mu}\xi_{\mu}),
\end{equation}
where $\xi^{\mu}$ is a local time-like Killing vector. Then we can
write the Unruh temperature as
\begin{equation}\label{vt}
T={\hbar\over 2\pi}e^\phi n^{\mu} \nabla_{\mu} \phi,
\end{equation}
where $n^{\mu}$ is a unit vector normal to the time-like surface
$^3B$ (please refer to the next section for the definition of
$^3B$).

The key assumptions of Verlinde's approach are the equipartition
theorem and the Tolman-Komar mass
\begin{eqnarray}
M&=&\frac{TN}{2}=\frac{TA}{2G\hbar},\label{energy}\\
M&=&2\int_V (T_{\mu\nu}-\frac{T}{2}g_{\mu\nu})u^{\mu}\xi^{\nu}dV
,\label{Tollman}
\end{eqnarray}
where $N=\frac{A}{G\hbar}$ is supposed to be the number of degrees
of freedom on the screen, $V$ is a spacelike hypersurface denoting
the space volume and $u^{\mu}$ is the unit vector normal to $V$. It
should be mentioned that the the equipartition rule of gravity
eq.(\ref{energy}) was first derived by Padmanabhan \cite{Pad1}.
Substituting eq.(\ref{vt}) into eq.(\ref{energy}), we derive
\cite{Verlinde}
\begin{eqnarray}\label{energy1}
 M=\frac{TN}{2}=\frac{TA}{2G\hbar}=\frac{1}{4\pi
 G}\int_V R_{\mu\nu}u^{\mu}\xi^{\nu}dV.
\end{eqnarray}
Equating eq.(\ref{Tollman}) and eq.(\ref{energy1}), we obtain the
Einstein equations in an integral form
\begin{eqnarray}\label{Vequation}
\int_V R_{\mu\nu}u^{\mu}\xi^{\nu}dV=8\pi G\int_V
(T_{\mu\nu}-\frac{T}{2}g_{\mu\nu})u^{\mu}\xi^{\nu}dV.
\end{eqnarray}
Note that $u^{\mu}, \xi^{\nu}$ are both time-like, so strictly
speaking, we can only derive the time-time (tt) component of
Einstein equations from eq.(\ref{Vequation}).

Now, let us turn to Jacobson's approach to derive the Einstein
equations on a null surface. For simplicity, we use a slightly
different method from Jacobson's initial approach. The key
hypothesis of Jacobson's approach is that the entropy is
proportional to area of the null surface
\begin{eqnarray}\label{Jentropy}
S= \eta A,
\end{eqnarray}
with $\eta$ a constant. Jacobson also assumes that the first law of
thermodynamics is satisfied on the null surface, thus we have
\begin{eqnarray}\label{Jfirstlaw}
\delta Q=T \delta S,
\end{eqnarray}
with $T$ the Unruh temperature. Note that on a null surface we have
\begin{eqnarray}\label{Jtem1}
T=\frac{\kappa}{2\pi}=\frac{1}{2\pi}k^{\mu}l^{\nu}\nabla_{\mu}\xi_{\nu},
\end{eqnarray}
where $\kappa$ is the surface gravity,
$k^{\mu}=\frac{dx^{\mu}}{d\lambda}$ is the tangent vector to the
horizon, $\lambda$ is the affine parameter, $l^{\nu}$ is an
auxiliary null vector with the property $l^{\mu}k_{\mu}=-1$. Using
$k^{\mu}\nabla_{\mu}k^{\nu}=0$ and $\xi^{\mu}=-\lambda \kappa
k^{\mu}$ on horizon, we can prove eq.(\ref{Jtem1}).

Substituting eqs.(\ref{Jentropy},\ref{Jtem1}) into
eq.(\ref{Jfirstlaw}), we can derive
\begin{eqnarray}\label{Jtem}
\delta Q&=&T \delta
S=\eta\frac{\kappa}{2\pi}(dA|^{\lambda+d\lambda}_{\lambda})\nonumber\\
&=&\frac{\eta}{2\pi}\int_{^2H}(k^{\mu}l^{\nu}\nabla_{\mu}\xi_{\nu}dA)|^{\lambda+d\lambda}_{\lambda}\nonumber\\
&=&\frac{\eta}{2\pi}\int_{H}k^{\mu}\nabla^{\nu}\nabla_{\mu}\xi_{\nu}dAd\lambda\nonumber\\
&=&\frac{\eta}{2\pi}\int_{H}k^{\mu}R_{\mu\nu}\xi^{\nu}dAd\lambda.\nonumber\\
\end{eqnarray}
In the above derivations, we have used Stokes's Theorem and the
identity $\nabla_{\nu}\nabla_{\mu}\xi^{\nu}=R_{\nu\mu}\xi^{\nu}$.

According to \cite{Jacobson}, all the heat flux passing through the
horizon is carried by matter
\begin{eqnarray}\label{flux}
\delta Q=\int_{H}T_{\mu\nu}k^{\mu}\xi^{\nu}dAd\lambda.\nonumber\\
\end{eqnarray}
In other words, the energy carried by matter becomes pure heat when
it passes through the horizon. That is because the energy of matter
inside the horizon is macroscopically unobservable. From
eqs.(\ref{Jtem},\ref{flux}) and $k^{\mu}\xi_{\mu}=0$, we can derive
\begin{eqnarray}\label{Jequation1}
R_{\mu\nu}+fg_{\mu\nu}=\frac{2\pi}{\eta}T_{\mu\nu}.
\end{eqnarray}
From the conservation of energy of matter
$\nabla_{\mu}T^{\mu\nu}=0$, we get $f=-\frac{R}{2}+\Lambda$. Set
$\eta=\frac{1}{4 G}$, we obtain the Einstein equations
\begin{eqnarray}\label{Jequation2}
R_{\mu\nu}-\frac{R}{2}g_{\mu\nu}+\Lambda g_{\mu\nu}=8\pi
GT_{\mu\nu}.
\end{eqnarray}
Strictly speaking, we can only derive the null-null (kk) component
of Einstein equations from eqs.(\ref{Jtem},\ref{flux}).

To end this section, let us compare the approaches of Verlinde,
Jacobson with ours of the next section in Table.\ref{table1}.
\begin{table}
\caption{comparisons between Verlinde's, Jacobson's and our methods}
\begin{center}
\label{table1}
\begin{tabular}{|c|c|c|c|}
  \hline\hline
   & assumption1 & assumption2 & Einstein equations\\
  \hline\hline
  Verlinde&Equipartition rule \ \ $M=NT/2~~~~~~$& Tolman-Komar mass& tt component~~~~~~~ \\
  \hline
 Jacobson &The first law \ \ \ $\delta Q=T\delta S$ & Entropy $S\sim A$ & kk(null) component\\
  \hline
  ours &The first law $\delta S=\beta (\delta E+p\delta A-\omega^a\delta J_a)$ & Surface stress tensor $\tau_{ij}$& All the components~\\
  \hline
\end{tabular}
\end{center}
\end{table}

\section{From holographic thermodynamics to vacuum Einstein equations}
In this section, we shall derive the vacuum Einstein equations from
the first law of thermodynamics on a time-like screen.

We use the notations of \cite{Brown1,Brown2} in this paper. Let us
make a brief review of these notations. $\Sigma_{t}$ is used to
denote a family of spacelike slices that foliate M. The boundary of
$\Sigma_{t}$ is B, which is supposed to be closed. The product of B
with segments of timelike world lines normal to $\Sigma_{t}$ at B is
denoted as $^{3}B$, the time-like three-boundary of M. The above
notations are depicted in fig.1.
\begin{figure}
\begin{center}
\setlength{\unitlength}{0.55cm}
\begin{picture}(8,8)
\thicklines \qbezier(3,2)(5,3)(7,2) \qbezier(3,2)(5,1)(7,2)
\put(3,2){\line(0,1){4}} \put(7,2){\line(0,1){4}}
\qbezier(3,6)(5,7)(7,6) \qbezier(3,6)(5,5)(7,6)
\put(4.5,5.8){$\Sigma_{t^{''}}$} \put(4.5,1.8){$\Sigma_{t^{'}}$}
\put(6.5,3.1){$^{3}B$} \qbezier(3,4.5)(5,5.5)(7,4.5)
\qbezier(3,4.5)(5,3.6)(7,4.5) \put(5.7,3.9){B}
\end{picture}
\end{center}
\caption{Spacetime M with its three-boundary consisting of the
spacelike hypersurface $\Sigma_{t^{'}}$, $\Sigma_{t^{''}}$ and
timelike hypersurface $^{3}B$, the spacelike two-boundary of
$\Sigma_{t}$ is B. } \label{fig1}
\end{figure}
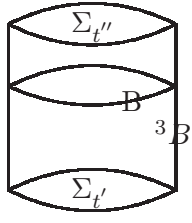

The metric and covariant derivative in $M$ are denoted by
$g_{\mu\nu}$ and $\nabla_{\mu}$, respectively. $n^{\mu}$ is the
outward pointing unit vector normal to $^{3}B$.
$\gamma_{\mu\nu}=g_{\mu\nu}-n_{\mu}n_{\nu}$ is the metric on $^3B$
with the corresponding covariant derivative $D_{\mu}$. The extrinsic
curvature on $^{3}B$ is defined as
\begin{equation}\label{theta}
\Theta_{\mu\nu}=-\gamma^{\alpha}_{\mu}\nabla_{\alpha}n_{\nu}.
\end{equation}
Let $x^{i}$ ($i=0,1,2$) be the intrinsic coordinates on $^{3}B$,
then we can rewrite the metric, covariant derivative and the
extrinsic curvature on $^3B$ as $\gamma_{ij}$, $D_i$ and
$\Theta_{ij}$ with only intrinsic three indexes. Let $u_{\mu}$ be
the future pointing unit vector normal to $\Sigma_{t}$, then the
metric and extrinsic curvature on $\Sigma$ can be defined as
$h_{\mu\nu}=g_{\mu\nu}+u_{\mu}u_{\nu}$ and $K_{\mu\nu}=-h_{\mu}^{\
\alpha}\nabla_{\alpha}u_{\nu}$, respectively. For simplicity, we
require that $(u\cdot n)|_{^{3}B}=0$ on the time-like screen as
\cite{Brown1,Brown2}, which means that the observer co-moving with
$^{3}B$ is static with respect to hypersurfaces $\Sigma_{t}$. With
the above restriction, the metric on $^{3}B$ can be decomposed as
\begin{eqnarray}\label{ADM}
\gamma_{ij}dx^idx^j=-N^2dt^2+\sigma_{ab}(dx^a+V^adt)(dx^b+V^bdt),
\end{eqnarray}
where $x^a$($a=1,2$) and $\sigma_{ab}$ are coordinates and metric on
$B$, respectively.

Let us begin to define various thermodynamic quantities on the
time-like screen $^{3}B$. Suppose that the stress tensor on the
time-like screen is $\tau^{ij}$. Now we do not assume any specific
form of $\tau^{ij}$. In fact, one of the main purposes of this paper
is to find a suitable expression for $\tau^{ij}$. Using the surface
stress tensor $\tau^{ij}$, we can define the energy density
$\varepsilon$, angular momentum density $j_a$ and spatial stress
$s^{ab}$ on the time-like screen as follows:
\begin{eqnarray}\label{Ejs}
&&\varepsilon=u_i u_j \tau^{ij},\\
&&j_a=-\sigma_{ai}u_j \tau^{ij},\\
&&s^{ab}=\sigma^a_i\sigma^b_j \tau^{ij},
\end{eqnarray}
where $u_i=(-N,0,0)$ is the speed of observer and $\sigma_{ai}$ is
the projection operator from $^{3}B$ to space-like two-boundary $B$
\cite{Brown1,Brown2}. As usual, we can define the energy $E$,
angular momentum $J_a$, angular velocity $\omega^a$ and pressure
tensor $p^{ab}$ as
\begin{eqnarray}\label{Ejs1}
&&E=\int_B d^2x \sqrt{\sigma}\varepsilon,\\
&&J_a=\int_B d^2x \sqrt{\sigma}j_a,\\
&&\omega^a=\frac{V^a}{N},\ \ \ p^{ab}=s^{ab},
\end{eqnarray}
where the pressure $p=\frac{1}{2}s^{ab}\sigma_{ab}$ if $B$ is
homogeneous and isotropic. The first law of thermodynamics on the
time-like screen is
\begin{eqnarray}\label{firstlaw}
\delta S=\beta (\delta E - \omega^a \delta J_a +p\delta A),
\end{eqnarray}
where $S$ is the entropy, $\beta=\frac{1}{T}$ is the inverse
temperature, and $A=\int_B dx^2 \sqrt{\sigma}$ is the area of the
screen. In general, since the pressure on $B$ is a tensor, we can
rewrite the first law in a more general form as
\begin{eqnarray}\label{firstlaw1}
\delta S=\beta\int_{B} d^2x[\delta (\sqrt{\sigma}\varepsilon) -
\frac{V^a}{N}\delta(\sqrt{\sigma}j_a)
+\frac{\sqrt{\sigma}}{2}s^{ab}\delta \sigma_{ab}].
\end{eqnarray}

Now we start to discuss the suitable definitions of $\beta$ on a
time-like screen. To get some insights for the definitions of
$\beta$, let us first consider a special case of static black hole
surrounded by a screen. Suppose that the inverse temperature of
black hoe is $\beta_0$, then the inverse temperature on the screen
is $\beta=N \beta_0$, where $N$ is the red-shift factor on the
screen. Note that for a static black hole, $\beta_0$ is the period
of the Euclidean time $\tau=i t$, $\beta_0=\oint d\tau$. So we get
$\beta=\oint d\tau N=i\int dt N$. Based on the above discussions, we
assume $\beta$ can be defined as
\begin{eqnarray}\label{temperature}
\beta=i\int dt N|_{B},
\end{eqnarray}
 with $t$ a pure imaginary number. For simplicity, we redefine $(it)$ as $t$.
 Now $t$ is a real number and the first law eq.(\ref{firstlaw1}) becomes
\begin{eqnarray}\label{firstlaw2}
\delta S=\int dt\int_{B} d^2x N[\delta (\sqrt{\sigma}\varepsilon) -
\frac{V^a}{N}\delta(\sqrt{\sigma}j_a)
+\frac{\sqrt{\sigma}}{2}s^{ab}\delta \sigma_{ab}].
\end{eqnarray}

Let us go on to discuss the suitable definition of stress tensor
$\tau^{ij}$ on a time-like screen. First, we aim to derive the field
equations of gravitation in the bulk from the first law of
thermodynamics on a screen. Due to the universality of gravity,
$\tau^{ij}$ must be independent of the details of matter fields.
Thus, $\tau^{ij}$ should be a pure geometric quantity. Second, we
require $\tau^{ij}$ to be a tensor on the screen so that the entropy
$S$ and gravitational field equations to be derived are independent
of the choices of coordinates on the screen. Third, note that on a
$1+2\ D$ screen, there are only three independent second-order
tensors: the metric $\gamma_{ij}$, Ricci curvature $\bar{R}_{ij}$
and extrinsic curvature $\Theta_{ij}$. For simplicity, we assume
that $\tau_{ij}$ is only linearly dependent of Ricci curvature and
extrinsic curvature:
\begin{eqnarray}\label{t}
\tau_{ij}=c_1 \bar{R}_{ij}+c_2\ \Theta_{ij}+f \gamma_{ij},
\end{eqnarray}
where $c_1, c_2$ are two constants and $f$ is a function to be
determined. It should be stressed that $\tau_{ij}$ must depend on
the extrinsic curvature $\Theta_{ij}$. Since the intrinsic geometry
of the screen and the bulk geometry are independent, if $\tau_{ij}$
depends only on the intrinsic geometry $\gamma_{ij},\ \bar{R}_{ij}$
of the screen, we could not get any information of the dynamics of
bulk geometry. One the other hand, the extrinsic curvature contains
both the information of bulk and screen. Thus, it is a natural
holographic bridge between the thermodynamics on the screen and the
dynamics of geometry in the bulk.

Now with the suitable hypothesis of inverse temperature
eq.(\ref{temperature}) and surface stress tensor eq.(\ref{t}), let
us find out what can we obtain from the first law of thermodynamics
eq.(\ref{firstlaw2}) on the holographic screen. Substituting
eq.(\ref{t}) into eq.(\ref{firstlaw2}), we can derive

\begin{eqnarray}\label{equation}
\delta S=&&\int_{^{3}B} dtd^2x N[\delta (\sqrt{\sigma}\varepsilon) -
\frac{V^a}{N}\delta(\sqrt{\sigma}j_a)
+\frac{\sqrt{\sigma}}{2}s^{ab}\delta \sigma_{ab}]\nonumber\\
=&& \delta S_0
+\frac{c_2}{2}\int_{M}d^4x\sqrt{-g}(R^{\mu\nu}-\frac{R}{2}g^{\mu\nu}+\Lambda
g^{\mu\nu})\delta
g_{\mu\nu}\nonumber\\
&-&\frac{c_2}{2}\int_{\Sigma}\sqrt{h}d^3x(Kh^{ij}-K^{ij})\delta
h_{ij}|^{t''}_{t'}\nonumber\\
&-&\int_{^3B}d^3x\sqrt{-\gamma}\delta(f+c_2\Theta+\frac{c_1\bar{R}}{2})
\end{eqnarray}
where $S_0$ is
\begin{eqnarray}\label{S0}
S_0=&&\frac{c_2}{2}\int_M d^4x\sqrt{-g}(R-2\Lambda)+c_2\int_{\Sigma}\sqrt{h}d^3xK|^{t''}_{t'}\nonumber\\
&-&c_2 \int_{^{3}B}d^3x\sqrt{-\gamma}t_{i}\Theta^{ij}\partial_{j}t
-c_1\int_{^{3}B}d^3x\sqrt{-\gamma}t_{i}\bar{R}^{ij}\partial_{j}t
\end{eqnarray}
with $t^{i}=Nu^{i}+V^{j}$ and $\partial_{i}t=-u_{i}/N$. Note that
the terms on the space-like boundary $\Sigma$ vanish when $M$ has a
topology $M=(disk)\times S^2$ with $\partial M=S^1\times S^2$. That
is because the manifolds considered here have a single boundary
$\partial M=^3 B$. In a more general case, we should consider the
terms on $\Sigma$. However, for simplicity, we shall consider the
terms on $\Sigma$ only in this section and ignore them in the
following sections.

Because the entropy $S$ is a total differential, the second terms of
the second line of eq.({\ref{equation}), the third and fourth lines
of eq.({\ref{equation}) must vanish. So, we can derive
\begin{eqnarray}\label{S}
&&S=S_0,\\
&&R_{\mu\nu}-\frac{R}{2}g_{\mu\nu}+\Lambda
g_{\mu\nu}=0,\\
&&\delta h_{ij}|_{\Sigma_{t'}}=\delta h_{ij}|_{\Sigma_{t''}}=0,\label{hij}\\
&&f=-c_1\frac{\bar{R}}{2}-c_2\Theta+c_3,
\end{eqnarray}
where $c_3$ is a constant and eq.(\ref{hij}) is the boundary
condition on $\Sigma$. Now from the first law of thermodynamics on
the screen, we have derived the vacuum Einstein equations, the
formula of entropy and the surface stress tensor
\begin{eqnarray}\label{tensor}
\tau_{ij}=c_2(\Theta_{ij}-\Theta
\gamma_{ij})+c_1(\bar{R}_{ij}-\frac{\bar{R}}{2}\gamma_{ij})+c_3\gamma_{ij}.
\end{eqnarray}

Notice that the first term of eq.(\ref{tensor}) is just the
Brown-York quasilocal stress tensor \cite{Brown1} if we set
$c_2=\frac{1}{8\pi G}$. Form eq.(\ref{equation}), it is easy to
observe that the last two terms of the surface stress tensor
eq.(\ref{tensor}) do not affect the bulk field equations. Below, we
shall prove that the last two terms do not affect the entropy
eq.(\ref{S0}) either. For simplicity, we focus on the stationary
spacetime. We only need to prove that the last term of eq.(\ref{S0})
vanishes for stationary spacetime. Substituting
$\partial_{i}t=-u_{i}/N$ and $\sqrt{-\gamma}=N\sqrt{\sigma}$ into
the last term of eq.(\ref{S0}), we can derive
\begin{eqnarray}\label{c2S}
&-&c_1
\int_{^{3}B}d^3x\sqrt{-\gamma}t_{i}\bar{R}^{ij}\partial_{j}t\nonumber\\
&=& c_1\int dt \int_{B}d^2x\sqrt{\sigma}t_{i}\bar{R}^{ij}u_j\nonumber\\
&=& c_1\int dt \int_{B}d^2x\sqrt{\sigma}u^{j}D^iD_{[j}t_{i]}\nonumber\\
&=& c_1\int dt \int_{\partial
B}dx\sqrt{\sigma}u^{j}m^iD_{[j}t_{i]}=0.
\end{eqnarray}
In the above derivations, we have used the fact that $t^{i}=(1,0,0)$
is a Killing vector in stationary spacetime. So we have
$D_{(i}t_{j)}=0$ and $\bar{R}_{ij}t^j=D_jD_it^j$. We also used the
property that $B$ has no boundary. Now we find that the last two
terms of the surface stress tensor eq.(\ref{tensor}) have no
relations with the entropy and the gravitational field equations. In
fact, they are related to the freedom to choose the zero point of
the energy according to \cite{Brown1}. We want to mention that we
can add extra terms $\tau^{e}_{ij}$ to eq.(\ref{tensor}) if we relax
the constraint that $\tau_{ij}$ depends on $\bar{R}_{ij}$ linearly.
A general form of such terms are as follows:
\begin{eqnarray}\label{terms}
\tau^e_{ij}=L \gamma_{ij}-2P_i^{lmn}\bar{R}_{jlmn}+4D^nD^mP_{inmj},
\end{eqnarray}
where $P^{ijkl}=\frac{\partial L}{\partial R_{ijkl}}$ with $L$ is a
general scalar function of $\gamma_{ij},\ \bar{R}_{ijkl}$. One can
prove that such terms eq.(\ref{terms}) do not affect the bulk field
equations and the entropy in stationary spacetime.

Now let us turn to study the entropy eq.({\ref{S0}). First, we want
to mention that if $(N,\  V^i,\  h_{ij})$ is a solution of Einstein
equations with time $t$, then $(-iN,\ -iV^i,\ h_{ij})$ is a solution
of Einstein equations with imaginary time $(i\ t)$. And all the
extensive quantities such as the entropy, energy, angular momentum
and the area remain invariant under such transformations
$(t\rightarrow i t,\ N\rightarrow -i N,\ V^i\rightarrow -i V^i,
h_{ij}\rightarrow h_{ij})$ \cite{Brown2}. So we can rewrite the
entropy as $S(t, N, V^j, h_{ij})=S(it, -iN, -i V^j, h_{ij})$ which
is consistent with eq.(4.9) of \cite{Brown2} in stationary
spacetime. According to \cite{Brown2}, we can rewrite eq.({\ref{S0})
as
\begin{eqnarray}\label{S1}
S_0=&&\int_M d^4x(P^{ij}\dot{h}_{ij}-N H-V^{i}H_{i})\nonumber\\
&+&\int_{^{3}H}d^3x\sqrt{\sigma}[n^i(\partial_i N)/(8\pi G)+2n_iV_jP^{ij}/\sqrt{h}]\nonumber\\
&-&c_1 \int_{^{3}B}d^3x\sqrt{-\gamma}t_{i}\bar{R}^{ij}\partial_{j}t,
\end{eqnarray}
where we have set $c_2=\frac{1}{8\pi G}$ and $^3H$ is the horizon
when there is a black hole contained in the screen.
$H=-\frac{\sqrt{h}}{8\pi G}G_{\mu\nu}u^{\mu}u^{\nu}$ and
$H_i=-\frac{\sqrt{h}}{8\pi G}G_{\mu\nu}u^{\mu}h^{\nu}_{\ i}$ are the
Hamiltonian and momentum constraints, respectively. For simplicity,
we focus on the stationary spacetime. Then, the first line and last
line of the above equation vanish due to stationarity and bulk field
equations. Notice that we have $N=V^i=0$, $\int dt=\beta_0$ and
$\beta_0n^i\partial_i N=2\pi$ on the horizon \cite{Brown2}. So we
can derive the entropy as
\begin{eqnarray}\label{S2}
S&=&S_0=\frac{1}{8\pi G}\int dt \int d^2x\sqrt{\sigma}n^i\partial_i N \nonumber\\
&=&\frac{1}{4}\int d^2x\sqrt{\sigma}=\frac{A_{H}}{4},
\end{eqnarray}
which is just the Bekenstein-Hawking entropy.

To summarize, we have derived the vacuum Einstein equations from the
first law of thermodynamics on a holographic time-like screen. We
also obtain the correct black hole entropy. With the hypothesis that
$\tau_{ij}$ depends on surface Ricci curvature and extrinsic
curvature linearly, we can prove that $\tau_{ij}$ is just the
Brown-York stress tensor (set $c_2=\frac{1}{8\pi G}$) plus terms
which do not affect the entropy (in stationary spacetime) and the
field equations of gravitation. We want to mention that for a more
general hypothesis of $\tau_{ij}$ one may derive other gravity
theories. However, we only investigate the simplest case in this
paper.

To end this section, let us study a simple example to help us
understand the above holographic derivations of Einstein equations.
For simplicity, we focus on a spherically symmetric static spacetime
with the metric
\begin{eqnarray}\label{example}
ds^2=-N^2dt^2+h^2dr^2+r^2d\Omega^2,
\end{eqnarray}
where $N$ and $h$ are functions of $r$ and some parameters such as
mass or entropy. And the screen is chosen to be the hypersurface at
fixed $r$. For simplicity, we suppose that the stress tensor on the
screen takes the form of eq.(\ref{tensor})
\begin{eqnarray}\label{t2}
\tau_{ij}=\frac{1}{8\pi G}\ (\Theta_{ij}-\Theta \gamma_{ij})+c_1
(\bar{R}_{ij}-\frac{\bar{R}}{2}\gamma_{ij})+c_3 \gamma_{ij},
\end{eqnarray}
where we have set $c_2=\frac{1}{8\pi G}$ for simplicity. Form
eq.(\ref{Ejs1}), we get
\begin{eqnarray}\label{Ep}
E&=&-\frac{r}{h}-4\pi r^2 c_3+4\pi c_1,\\
p&=&\frac{1}{8\pi}(\frac{N'}{Nh}+\frac{1}{rh})+c_3.
\end{eqnarray}
We suppose that $N, \ h$ are functions of the area $A=4\pi r^{2}$
and the entropy $S$ ($A$ and $S$ are two independent thermodynamic
quantities on the screen), then the first law of thermodynamics on
the screen becomes
\begin{eqnarray}\label{law}
dS&=&\beta(dE+pdA)\nonumber\\
&=&\beta[\frac{1}{8\pi h}(\frac{h'}{h}+\frac{N'}{N})\
dA+\frac{r}{h^2}(\partial_{S}h) \ dS]
\end{eqnarray}
Since $S$ is a total differential, we have
\begin{eqnarray}
\frac{h'}{h}+\frac{N'}{N}=0\label{equ1},\\
\beta\frac{r}{h^2}(\partial_{S}h)=1.\label{equ2}
\end{eqnarray}
It is interesting to note that eqs.(\ref{law}-\ref{equ2}) are
independent of $c_1$ and $c_3$, which is consistent with our above
conclusion that the last two terms of eq.(\ref{tensor}) do not
affect the entropy and bulk field equations. From eq.(\ref{equ1}),
one can easily get $h=c/N$. By redefining $t$, we can set $c=1$. So
we get
\begin{eqnarray}\label{NN}
h=\frac{1}{N}.
\end{eqnarray}
Now we suppose there is a horizon at $r_H$ inside the screen.
Following the standard procedure, to avoid the conical singularity
at the horizon, we can derive the temperature of the horizon as
$T_H(S)=\frac{NN'}{2\pi}|_{r=r_H}$. Then the inverse temperature on
the screen is
\begin{eqnarray}\label{TT}
\beta=\frac{N}{T_H(S)}=N(\frac{2\pi}{NN'}|_{r=r_H})
\end{eqnarray}
Substituting eq.(\ref{NN}) and eq.(\ref{TT}) into eq.(\ref{equ2}),
we get
\begin{eqnarray}\label{fu}
\frac{\partial N^2}{\partial S}=-2\frac{T_H(S)}{r},
\end{eqnarray}
from which we can derive $N^2=c-\frac{2f(S)}{r}$ with $df=T_{H}dS$.
By redefining $t$, we can again set $c=1$. Thus, we get
\begin{eqnarray}\label{fuu}
N^2=1-\frac{2f(S)}{r}.
\end{eqnarray}
Applying $T_H(S)=\frac{NN'}{2\pi}|_{r=r_H=2f(S)}$ together with
$df(S)=T_HdS$ and eq.(\ref{fuu}), we can derive
\begin{eqnarray}\label{en1}
T_H=\frac{1}{8\pi f(S)}, \ S=\frac{1}{16\pi T_{H}^2}=\frac{A_H}{4}.
\end{eqnarray}
Now we have derived the correct spherically symmetric static
solution of vacuum Einstein equations and black hole entropy:
\begin{eqnarray}\label{solution}
N^2=1/h^2=1-\frac{\sqrt{S/\pi}}{r},\ \ S=\frac{A_H}{4}.
\end{eqnarray}

\section{From holographic thermodynamics to Einstein equations with matter}
In this section, we generalize our holographic program to the case
with matter fields. We assume the ``Fine first law of thermodynamics"
on the screen, with which we mean that there is a corresponding ``fine
matter term" added to the first law for every special kind of matter
field. By assuming the ``Fine first law of thermodynamics", we can
not only derive the Einstein equations with matter but also the
matter field equations. So it is a very powerful holographic
program. The key point of this holographic program is to search for
a suitable form of the ``fine matter terms" on the screen. We take
scalar field and electromagnetic field as examples below.

Let us firstly study the case of scalar field $\phi$. We suppose the
``Fine first law of thermodynamics" on the screen is
\begin{eqnarray}\label{Finefirstlaw}
\delta S=\beta (\delta E - \omega^a \delta J_a +p\delta A+F_\phi
\delta \phi),
\end{eqnarray}
where
\begin{eqnarray}\label{J}
F_\phi=-\int_Bd^2x\sqrt{\sigma}n^{\mu}\partial_\mu \phi.
\end{eqnarray}
According to \cite{Creighton}, an extensive variable is a function
of the phase space coordinates on $B$ only. Thus, $\phi$ is an
extensive variable and the first law of thermodynamics
eq.(\ref{Finefirstlaw}) includes only variation of extensive
variables. $F_\phi$ is designed so that we can derive the usual
scalar field equation in the bulk. We assume the surface stress
tensor take the form of eq.(\ref{t}) as in the above section.
Following a similar program, we can derive
\begin{eqnarray}\label{dS1}
\delta S&=&\int_{^{3}B} dtd^2x N[\delta (\sqrt{\sigma}\varepsilon) -
\frac{V^a}{N}\delta(\sqrt{\sigma}j_a)
+\frac{\sqrt{\sigma}}{2}s^{ab}\delta \sigma_{ab}-\sqrt{\sigma}n^{\mu}\partial_\mu \phi\delta\phi]\nonumber\\
&=& \delta S_1
+\frac{c_2}{2}\int_{M}d^4x\sqrt{-g}(R^{\mu\nu}-\frac{R}{2}g^{\mu\nu}+\Lambda
g^{\mu\nu}-\frac{1}{c_2}T^{\mu\nu})\delta
g_{\mu\nu}\nonumber\\
&&-\int_{^3B}d^3x\sqrt{-\gamma}\delta(f+c_2\Theta+\frac{c_1\bar{R}}{2})-\int_Md^4x\sqrt{-g}[\Box\phi-\frac{\partial
V(\phi)}{\partial\phi}]\delta\phi,
\end{eqnarray}
where $S_1$ and $T^{\mu\nu}$ are
\begin{eqnarray}\label{S1T}
&&S_1=S_0+\int_Md^4x\sqrt{-g}[-\frac{1}{2}\partial_\mu\phi\partial^{\mu}\phi-V(\phi)],\label{S1T}\\
&&T^{\mu\nu}=\frac{2}{\sqrt{-g}}\frac{\delta (S_1-S_0)}{\delta
g_{\mu\nu}}\label{S1TTT}.
\end{eqnarray}
Since the entropy is a a total differential, from eq.(\ref{dS1}) we
can derive the Einstein equations, scalar field equation, the
entropy and the surface stress tensor as follows:
\begin{eqnarray}\label{S}
&&R_{\mu\nu}-\frac{R}{2}g_{\mu\nu}+\Lambda g_{\mu\nu}=8\pi G T_{\mu\nu},\\
&&\Box\phi-\frac{\partial V(\phi)}{\partial\phi}=0,\ \ S=S_1,\\
&&\tau_{ij}=\frac{1}{8\pi G}(\Theta_{ij}-\Theta
\gamma_{ij})+c_1(\bar{R}_{ij}-\frac{\bar{R}}{2}\gamma_{ij})+c_3\gamma_{ij},
\end{eqnarray}
where we have set $c_2=\frac{1}{8\pi G}$. Following the same method
of Sec.~3, one can easily prove that the entropy of stationary black
hole with scalar field is also $S=\frac{A_H}{4}$.
\begin{eqnarray}\label{scalarentropy}
S=&&\int_M d^4x(P^{ij}\dot{h}_{ij}+P_{\phi}\dot{\phi}-N H-V^{i}H_{i})\nonumber\\
&+&\int_{^{3}H}d^3x\sqrt{\sigma}n^i(\partial_i N)/(8\pi G)\nonumber\\
\end{eqnarray}
where $H=-\frac{\sqrt{h}}{8\pi G}(G_{\mu\nu}-8\pi G
T_{\mu\nu})u^{\mu}u^{\nu}$ and $H_i=-\frac{\sqrt{h}}{8\pi
G}(G_{\mu\nu}-8\pi G T_{\mu\nu})u^{\mu}h^{\nu}_{\ i}$ are the
Hamiltonian and momentum constraints. Note that, in general, we have
$\sqrt{-g}n^{\mu}\partial_{\mu}\phi\rightarrow 0$ as $r\rightarrow
\infty$. Thus, the first law of thermodynamics
eq.(\ref{Finefirstlaw}) reduces to the usual form $\delta S=\beta
(\delta E - \omega^a \delta J_a +p\delta A)$ on a screen placed at
infinity. And the ``fine matter term" $F_\phi \delta\phi$ only
becomes important on the screen at finite radius.

Finally, we want to clarify the confusion readers may have in our
above derivations. It seems that there are arbitrary terms of
potential energy $V(\phi)$ and cosmological constant $\Lambda$ in
our above derivations. Thus, the same first law of thermodynamics on
the screen appears to correspond to many different bulk field
equations with different $V(\phi)$ and $\Lambda$. This is however
not the case. Notice that the entropy eq.(\ref{S1T}) in the first
law is different for different choice of $V(\phi)$ and $\Lambda$. So
one specific first law of thermodynamics on the screen corresponds
to the unique bulk field equations. In fact, similar to the integral
constant, $\sqrt{-g}V(\phi)$ and $\sqrt{-g}\Lambda$ can be regarded
as the ``integral constant functions" which do not affect the
variational forms of the first law.

Now we turn to discuss the case of electromagnetic field. We develop
a two-step program. In the first step, we focus on the derivations
of bulk field equations, so we do not need all the details of the
first law of thermodynamics on the screen. Instead, we leave a total
differential surface term $S_A$ free which does not affect the bulk
field equations. In the second step, we choose $S_A$ carefully so
that the first law of thermodynamics on the screen includes only
variation of extensive variables. And as a result, we can derive the
entropy of the system.

First, we suppose the ``Fine first law of thermodynamics" on the
screen is
\begin{eqnarray}\label{Finefirstlaw1}
\delta S=\beta (\delta E - \omega^a \delta J_a +p\delta A+F_A^{\mu}
\delta A_{\mu})+\delta S_A,
\end{eqnarray}
where $S_A$ is a total differential term which do not affect the
bulk field equations and $F_A^{\mu}$ is designed as
\begin{eqnarray}\label{J1}
F_A^{\mu}=-\int_Bd^2x\sqrt{\sigma}n_{\nu}F^{\nu\mu},
\end{eqnarray}
in order to derive the Maxwell's equations in the bulk. We assume
the surface stress tensor eq.(\ref{t}) as before. Following the same
program, we can derive
\begin{eqnarray}\label{dS2}
\delta S&=&\int_{^{3}B} dtd^2x N[\delta (\sqrt{\sigma}\varepsilon) -
\frac{V^a}{N}\delta(\sqrt{\sigma}j_a)
+\frac{\sqrt{\sigma}}{2}s^{ab}\delta \sigma_{ab}-\sqrt{\sigma}n_{\nu}F^{\nu\mu}\delta A_{\mu}]+\delta S_A\nonumber\\
&=& \delta S_2
+\frac{c_2}{2}\int_{M}d^4x\sqrt{-g}(R^{\mu\nu}-\frac{R}{2}g^{\mu\nu}+\Lambda
g^{\mu\nu}-\frac{1}{c_2}T^{\mu\nu})\delta
g_{\mu\nu}\nonumber\\
&&-\int_{^3B}d^3x\sqrt{-\gamma}\delta(f+c_2\Theta+\frac{c_1\bar{R}}{2})-\int_Md^4x\sqrt{-g}(\nabla_{\mu}F^{\mu\nu})\delta
A_{\nu},
\end{eqnarray}
where $S_2$ and $T^{\mu\nu}$ are
\begin{eqnarray}\label{S1T1}
&&S_2=S_0+S_A-\frac{1}{4}\int_Md^4x\sqrt{-g}F_{\mu\nu}F^{\mu\nu},\nonumber\\
&&T^{\mu\nu}=\frac{2}{\sqrt{-g}}\frac{\delta (S_2-S_0-S_A)}{\delta
g_{\mu\nu}}.
\end{eqnarray}
Because the entropy is a a total differential, from eq.(\ref{dS2})
we can derive the Einstein equations, Maxwell's equations and the
surface stress tensor as follows:
\begin{eqnarray}\label{S1}
&&R_{\mu\nu}-\frac{R}{2}g_{\mu\nu}+\Lambda g_{\mu\nu}=8\pi G T_{\mu\nu},\ \ \ \ \nabla_{\mu}F^{\mu\nu}=0,\nonumber\\
&&\tau_{ij}=\frac{1}{8\pi G}(\Theta_{ij}-\Theta
\gamma_{ij})+c_1(\bar{R}_{ij}-\frac{\bar{R}}{2}\gamma_{ij})+c_3\gamma_{ij},
\end{eqnarray}
where we have set $c_2=\frac{1}{8\pi G}$ as before.

Second, let us now study $S_A$ carefully so that we can express the
first law eq.(\ref{Finefirstlaw1}) with only variation of the extensive
variables obviously. Let us firstly rewrite the electromagnetic
terms of eq.(\ref{Finefirstlaw1}) as follows
\begin{eqnarray}\label{etrem}
&&\delta S_A-\int_{^{3}B}d^3xN\sqrt{\sigma}n_{\nu}F^{\nu i}\delta A_{i},\nonumber\\
&=&\delta S_A+\int_{^{3}B}d^3x\sqrt{\sigma}\{n_{\nu}F^{\nu
i}u_i[\delta (Nu^{j}A_{j})+\hat{A}_a\delta V^a]-Nn_{\nu}F^{\nu
a}\delta \hat{A}_a \}, \nonumber\\
&=&\delta S_A+\int_{^{3}B}d^3x\{-\tilde{Q}\delta
(N\Phi)+\tilde{J}_{a}\delta V^a]+NF^{a}\delta \hat{A}_a\},
\end{eqnarray}
where we have defined the electric potential $\Phi=-u^jA_j$, charge
density $\tilde{Q}=\sqrt{\sigma}n_{\mu}F^{\mu\nu}u_{\nu}$ and
electric momentum density $\tilde{J}_a=\tilde{Q}\hat{A}_a$ with
$\hat{A}_a=\sigma_a^{i}A_i$ and $F^a=-\sqrt{\sigma}n_{\nu}F^{\nu
a}$. In the above derivations, we have also used the formula $\delta
A_i=N^{-1}u_i[\delta (N\Phi)-\hat{A}_a\delta
V^{a}]+\sigma_i^a\delta\hat{A}_a$. Since the  extensive variables are
functions of only the phase space coordinates on $B$
\cite{Creighton}, $\tilde{J}_a,\ \tilde{Q}, \hat{A}_a,\ \sigma_{ab}$
are all the extensive variables. Conversely, $N,\ V^a,\ \Phi$ are the
intensive variables. Now it is clear by choosing $S_A$ as
\begin{eqnarray}\label{SA}
S_A=\int_{^{3}B}d^3x\{\tilde{Q}N\Phi-\tilde{J}_{a}V^a\},
\end{eqnarray}
the the first law of thermodynamics contains only variation of
extensive variables:
\begin{eqnarray}\label{dS3}
\delta &S&=\beta[\delta E -\omega^a\delta(J_a+\tilde{J}_a)+p\ dA+\Phi\delta Q+F^a\delta \hat{A}_a]\nonumber\\
&=&\int_{^{3}B} dtd^2x N[\delta (\sqrt{\sigma}\varepsilon) -
\frac{V^a}{N}\delta(\sqrt{\sigma}(j_a+\tilde{j}_a))
+\frac{\sqrt{\sigma}}{2}s^{ab}\delta \sigma_{ab}+\Phi\delta
\tilde{Q}-\sqrt{\sigma}n_{\nu}F^{\nu a}\delta
\hat{A}_{a}].\nonumber\\
\end{eqnarray}
Now we get the expression of the entropy
\begin{eqnarray}\label{S3}
S=S_0-\frac{1}{4}\int_Md^4x\sqrt{-g}F_{\mu\nu}F^{\mu\nu}+\int_{^{3}B}d^3x\{\tilde{Q}N\Phi-\tilde{J}_{a}V^a\}.
\end{eqnarray}
Following the same method, one can prove $S=\frac{A_H}{4}$ for
stationary charged black holes. It is interesting that for
Reissner-Nordstrom black hole, the first law reduces to the usual
form
\begin{eqnarray}\label{dS5}
\delta S=\beta[\delta E +p\ \delta A+\Phi\delta Q]\nonumber
\end{eqnarray}
on a time-like screen.

Now we have derived the Einstein equations, scalar field equation as
well as Maxwell's equations from the ``Fine first law of
thermodynamics" on a holographic screen. One can obtain the field
equations of other matter in a similar program. However, if we do
not care about the details of matter fields, we can simplify the
above derivations and develop a more universal holographic program.
Consider a uncharged system surrounded by a screen at $r\rightarrow
\infty$, all the matter fields vanish quickly enough as $r$
approaches infinity so that the ``Fine matter terms" become less
important and in the leading order the first law of thermodynamics
takes the universal form
\begin{eqnarray}\label{abirdlaw}
\delta S=\beta (\delta E - \omega^a \delta J_a +p\delta A)+...
\end{eqnarray}
where $...$ means next leading order. Suppose the surface stress
tensor eq.(\ref{t}) as before, we can derive
\begin{eqnarray}\label{abirddS}
\delta S=&&\int_{^{3}B} dtd^2x N[\delta (\sqrt{\sigma}\varepsilon) -
\frac{V^a}{N}\delta(\sqrt{\sigma}j_a)
+\frac{\sqrt{\sigma}}{2}s^{ab}\delta \sigma_{ab}]+...\nonumber\\
=&& \delta S_3
+\frac{c_2}{2}\int_{M}d^4x\sqrt{-g}(R^{\mu\nu}-\frac{R}{2}g^{\mu\nu}+\Lambda
g^{\mu\nu}-\frac{1}{c_2}T^{\mu\nu})\delta
g_{\mu\nu}\nonumber\\
&-&\int_{^3B}d^3x\sqrt{-\gamma}\delta(f+c_2\Theta+\frac{c_1\bar{R}}{2})-\frac{\delta
S_M(g_{\mu\nu},\Psi)}{\delta \Psi}\delta \Psi \ ...
\end{eqnarray}
where $S_3=S_0+S_M$ with $S_M$ is the action of matter fields
$\Psi$. From the above equation, we can derive the Einstein
equations, matter field equations, the entropy in the leading order
and the stress tensor:
\begin{eqnarray}\label{S1}
&&R_{\mu\nu}-\frac{R}{2}g_{\mu\nu}+\Lambda g_{\mu\nu}=8\pi G T_{\mu\nu},\nonumber\\
&&\frac{\delta
S_M}{\delta \Psi}=0,\ \ \ S=S_0+S_M+... \nonumber\\
&&\tau_{ij}=\frac{1}{8\pi G}(\Theta_{ij}-\Theta
\gamma_{ij})+c_1(\bar{R}_{ij}-\frac{\bar{R}}{2}\gamma_{ij})+c_3\gamma_{ij}.
\end{eqnarray}

\section{Conclusions}

In this paper, we have derived the Einstein equations and the black hole
entropy from the first law of thermodynamics on a holographic
time-like screen. Based on the reasonable hypothesis that the
surface stress tensor depends on extrinsic curvature and surface
Ricci curvature linearly, we prove that the stress tensor on the
screen is just the Brown-York stress tensor plus terms which do not
affect the gravitational field equation and the entropy of the
system. Applying a generalized ``fine first law of thermodynamics" or
the usual first law of thermodynamics on the holographic screen, we
can also derive the matter field equations. It is interesting to
generalize our holographic approach to the case of higher derivative
gravity. For example, if we suppose that the surface stress tensor
takes a more general form
\begin{eqnarray}\label{fR}
\tau_{ij}=F'(R)\ \Theta_{ij}+f \gamma_{ij}+c_1 \bar{R}_{ij},
\end{eqnarray}
where $R$ is the Ricci scalar in the bulk and $F'(R)=\partial
F(R)/\partial R$, one may obtain the F(R) gravity from the first law
of thermodynamics on the screen (one may need a extra `fine matter
term' for $\phi=F'(R)$ in view of the equivalence between the
scalar-tensor theory and F(R) gravity). Besides, it is also
interesting to investigate our holographic approach in quantum
level. For example, how to calculate the quantum corrections for
black hole entropy, how to establish the quantum theory of matter
fields and even the gravity in our holographic thermodynamics
approach... We hope to address these issues in the following works.

To end this paper, we want to stress that, in certain sense, our
holographic thermodynamics approach is equivalent to the action
principle: they can be used to derive the same bulk field equations.
However, they have several significant differences. First, they have
different physical origins: our holographic approach is based on the
first law of thermodynamics, or in other words, the conservation of
energy on the screen; while the action principle is based on the
principle of least action that the path taken by a particle is the
one with least (extreme) action. Second, the action principle needs
suitable boundary conditions in order to have a good definition of
the variations of action. On the contrary, our holographic approach
does not need such boundary conditions (at least on $^3B$). Last but
not least, our holographic approach depends on heavily the existence
of gravity. One can not write a non-degenerate first law of
thermodynamics on the screen if there is no gravitational
interaction in the bulk. While there is no such constraint for the
action principle: one can easily write a action of matter field
without gravity. It seems that the action principle is more
universal than our approach. This is however not the case. Taking
into account the fact that all of the matter fields have
gravitational interaction, one has to always consider the matter and
gravity interaction together for every physical process. So the
third point above is an advantage rather than weakness of our
holographic approach: we predict the existence of gravity in certain
sense just as the string theory predicts supersymmetry and the extra
dimensions.

\section*{Acknowledgements}

R. X. Miao would like to thank Y. Tian, X. N. Wu and H. B. Zhang for
useful discussions. This research was supported by a NSFC grant
No.10535060/A050207, a NSFC grant No.10975172, a NSFC group grant
No.10821504 and Ministry of Science and Technology 973 program under
grant No.2007CB815401.

\end{document}